\documentclass[twocolumn,showpacs,preprintnumbers,amsmath,amssymb,prb]{revtex4}
\usepackage{amsmath,amsfonts,bm,graphicx}
\usepackage{color}

\begin{document}

\title{Band gaps in graphene via periodic electrostatic gating}

\author{Jesper Goor Pedersen$^1$ and Thomas Garm Pedersen$^{1,2}$}
\affiliation{$^1$Department of Physics and Nanotechnology
             Aalborg University, Skjernvej 4A
             DK-9220 Aalborg East, Denmark\\
             $^2$Center for Nanostructured Graphene (CNG), Aalborg University, DK-9220 Aalborg East, Denmark
             }
\date{\today}

\begin{abstract}
Much attention has been focused on ways of rendering graphene semiconducting.
We study periodically gated graphene in a tight-binding model and find that, contrary to predictions based
on the Dirac equation, it is possible to open a band gap at the Fermi level using electrostatic gating
of graphene. However, comparing to other methods of periodically modulating graphene, namely perforated graphene structures,
we find that the resulting band gap is significantly smaller. We discuss the intricate dependence of the band gap
on the magnitude of the gate potential as well as the exact geometry of the edge of the gate region. 
The role of the overlap of the eigenstates with the gate region is elucidated.
Considering more realistic gate potentials, 
we find that introducing smoothing in the potential distribution, even over a range of little more than a single carbon atom, reduces the attainable band gap significantly.
\end{abstract}


\maketitle

\section{Introduction}
While graphene\cite{Novoselov2004} has proven to be a remarkable material, with electronic properties that are interesting from a 
fundamental\cite{Neto2009,Novoselov2005} as well as a technological viewpoint,\cite{Geim2007,Geim2009} 
the absence of a band gap severely limits its possible applications.
Several methods have been proposed for opening a gap in graphene. Relying on quantum confinement effects, the most immediate
way of making graphene semiconducting is by reducing the dimensionality by cutting graphene into narrow ribbons. Such so-called
graphene nanoribbons (GNRs) have band gaps that in general scale inversely with the width of the GNR, but which are very 
sensitive to the exact geometry of the edge of the ribbon.\cite{Nakada1996,Brey2006,Son2006}
Related to these ideas, periodically perforated graphene, termed graphene antidot lattices, effectively result in a
network of ribbons, and has been shown to be an efficient way of inducing an appreciable band gap in graphene.
\cite{Pedersen2008a} This idea has been successfully applied to fabricate simple graphene-based semiconductor devices.\cite{Bai2010,Kim2010}
Modifying graphene via adsorption of hydrogen presents another route towards opening a gap in graphene,
with fully hydrogenated graphene exhibiting a band gap of several electron volts,\cite{Sofo2007,Elias2009} while
patterned hydrogen adsorption yields band structures resembling those of graphene antidot lattices, with 
reported band gaps of at least $450$~meV.\cite{Balog2010}

The prospect of opening a band gap in graphene via electrostatic gating is intriguing, since it would allow for
switching between semi-metallic and semiconducting behavior and to dynamically alter the band gap to fit specific applications.
This makes it significantly more flexible than proposals relying on structural modification of graphene.
However, a linearization of the tight-binding Hamiltonian of graphene, resulting in the now widely studied Dirac equation (DE) 
of graphene,\cite{Semenoff1984,Neto2009} suggests that the Dirac fermions of graphene cannot be confined
by electrostatic gating, due to the phenomenon of Klein tunneling.\cite{Novoselov2006,Beenakker2008} 
Thus, while periodic gating of usual semiconductor heterostructures such as, e.g., GaAs quantum wells, does induce gaps in the 
dispersion relation,\cite{Pedersen2008} 
previous theoretical studies have indicated that band gaps are induced for neither
one-dimensional \cite{Barbier2008, Barbier2009} nor two-dimensional\cite{Park2008} periodic gating of graphene.

These studies have taken as their starting point the Dirac model of graphene, which is a low-energy continuum model, ignoring
atomistic details.
Here, we instead use a more accurate tight-binding (TB) model to study periodically gated graphene. Contrary to predictions of 
continuum (Dirac) models, the TB model suggests that it is indeed possible to open a band gap in graphene via periodic gating.
The aim of this paper is two-fold: (i) To compare periodically gated graphene with graphene antidot lattices. In doing so
we will illustrate that, contrary to what may be expected from the Dirac equation, a sufficiently large scalar 
potential, i.e., not necessarily a mass term, 
yields a band structure that is highly similar to that of perforated graphene structures;
(ii) to serve as a feasibility study of periodic gating as a means of inducing a band gap in graphene.
To this end, we will illustrate and discuss the non-trivial dependence of the band gap on the gate potential, as well 
as the intricate relation between band gap and the edge geometry of the gated region. These results will also serve
to illustrate some of the key differences between graphene and ordinary two-dimensional electron gases. 
While, initially, the potential will be modeled as a simple step function, we will show below that introducing 
smoothing in the potential distribution severely reduces the attainable band gap. 

Continuum and atomistic models of periodically gated graphene have previously been compared in Ref.~\onlinecite{Zhang2010}.
That study, however, focused on a single value of the potential strength and only considered structures that are 
rotated $30^\circ$ compared to the ones of the present work and, therefore, do not necessarily display any band gap even for 
perforated structures.\cite{Petersen2011}  Moreover, in this work we examine in detail the non-trivial dependence of the band gap on the magnitude of 
the potential and we consider more realistic, smooth potential profiles. Finally, we elucidate the intricate dependence on the precise edge 
geometry and show how the energy gap correlates with the gate region overlap of electron and hole states.

\section{Models}
\begin{figure}
\begin{center}
\includegraphics[width=\linewidth]{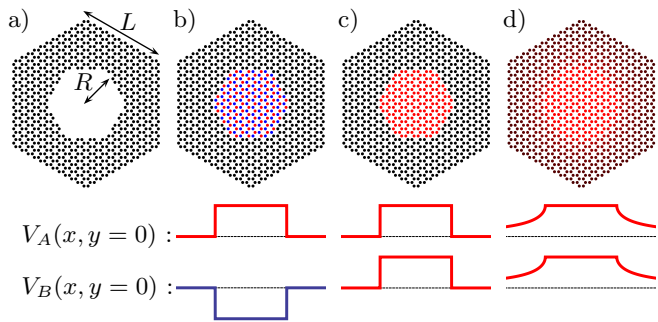}
\caption{(Color online) Unit cells used in the calculations for the $\{12,5\}$ lattice.
(a) Perforated graphene sheet, with carbon atoms removed in the region of the antidot.
(b) Staggered potential (mass term) in the antidot region. The color indicates the sign of the 
on-site energies.
(c) Constant gate potential in the antidot region.
(d) Gate potential modeled via Eq.~(\ref{eq:potential}), assuming the gate is directly below the graphene
sheet, with no insulating layer in-between.
The lower panel illustrates the potential of each model on the separate $A$ and $B$ sublattices.
}
\label{fig:geoms}
\end{center}
\end{figure}
In Fig.~\ref{fig:geoms} we illustrate the graphene structures that we will consider in this article. 
We consider only superlattices with triangular symmetry, as shown in the figure. An important decision lies
in the choice of the angle between the basis vectors of the superlattice and the carbon-carbon bonds in graphene.
In particular, if the superlattice basis vectors are rotated $30^\circ$ compared to the carbon-carbon bonds (such as in Ref.~\onlinecite{Zhang2010}), Clar sextet theory predicts that
perforated graphene structures only exhibit significant band gaps for every third value of the side length of
the hexagonal unit cell.\cite{Petersen2011} 
In contrast to this, perforated graphene structures with basis vectors parallel to the
carbon-carbon bonds always have band gaps.
We choose to focus in this paper on the latter geometries, in order to ensure that the superlattice symmetry
in itself does not prohibit the emergence of a band gap.

We characterize a given structure by 
$\{L,R\}$, where $L$ denotes the side length of the hexagonal unit cell, while $R$ is the radius of the
central region, both in units of the graphene lattice constant, as illustrated in Fig.~\ref{fig:geoms}. 
In these units, $L$ also corresponds to the number of benzene rings along each edge of the unit cell.
Note that the exact geometry of the edge of the central region differs greatly depending on the 
radius $R$. Below, we discuss in detail the crucial dependence of the results on the edge geometry.
We will consider four distinct ways of periodically modifying graphene:
(a) Perforated graphene (graphene antidot lattices), with carbon atoms removed from the central region,
(b) a periodic mass term, non-zero only in the central region, and
(c) periodically gated graphene, with a constant gate potential within the central region and a vanishing potential
outside.
Furthermore, to discuss the feasibility of realizing gapped graphene via periodic gating, we will also consider
(d) periodically gated graphene, with a more realistic model of the spatial dependence of the gate potential, obtained
from a solution to the Laplace equation.
Focus will be on periodically gated graphene, with the other forms of modulation included for comparison only.

To illustrate the dependence of the results on the exact edge of the gate or mass region, we will use a Dirac model as
well as a more accurate tight-binding treatment, in which the atomistic details of the structures are included.
We find significant discrepancies between these two methods, quantitatively as
well as qualitatively. In particular, we will show that the DE does not predict a band gap opening for periodic gating, which is
present in the TB results. In what follows, we briefly describe the two models.
In the continuum model of the problem, we employ the Dirac Hamiltonian
\begin{equation}
H_\mathrm{D} = \left[
\begin{array}{cc}
\Delta(x,y) & v_F\left(\hat{p}_x-i\hat{p}_y\right) \\
v_F\left(\hat{p}_x+i\hat{p}_y\right) & \pm\Delta(x,y)
\end{array}
\right],\label{eq:DE}
\end{equation}
where $v_F\simeq 10^6$~m/s is the Fermi velocity, while $\Delta(x,y)$ denotes the gate potential or mass term. 
Here, the $+$ ($-$) is used when modeling a gate potential (mass term). 
Imposing periodic Bloch boundary conditions at the edge of the unit cell, we solve the problem in a plane-wave spinor basis, 
$\left<\mathbf{r}|A_\mathbf{G}\right> = 
\bigl(\begin{smallmatrix} 1 \\0 \end{smallmatrix}\bigr)
e^{i(\mathbf{G}+\mathbf{k})\cdot\mathbf{r}}$
and
$\left<\mathbf{r}|B_\mathbf{G}\right> = 
\bigl(\begin{smallmatrix} 0 \\1 \end{smallmatrix}\bigr)
e^{i(\mathbf{G}+\mathbf{k})\cdot\mathbf{r}}$, with $\mathbf{k}$ the Bloch wave vector and $\mathbf{G}$ the reciprocal lattice 
vectors. We take $\Delta(\mathbf{r})=V_0\Theta(R-r)$,
with $\Theta(r)$ the Heaviside step function, yielding
$\Delta(\mathbf{G}) = 2\pi R V_0 J_1(GR)/(GA)$, where $A$ is the unit cell area while
$J_1(x)$ is the Bessel function of the first kind. A total of $1058$ plane-wave spinors were included in the calculations,
to ensure convergence of the results.

In the tight-binding model we include only nearest-neighbor coupling between $\pi$ orbitals, parametrized
via the hopping term $-t$, with $t=3$~eV. We ignore the overlap between neighboring $\pi$ orbitals, assuming that our basis
is orthogonal, and set the on-site energy of the $\pi$ orbitals to zero. This parametrization accurately reproduces
the Fermi velocity of graphene, and is also in quantitative agreement with density functional theory when applied
to perforated graphene structures.\cite{Furst2009} 
For periodically gated graphene, 
we set the diagonal terms of the Hamiltonian equal to the gate potential. In the case of
a mass term, the diagonal terms become $\pm V_0$, with the sign depending on which sublattice the carbon atom resides on.
For perforated graphene, atoms are removed entirely in the region of the hole, ensuring that no dangling bonds
are created.
While including next-nearest neighbor coupling, as well as 
taking into account the non-orthogonality of the basis set, will change our results quantitatively, we
expect the overall trends and the conclusions to remain the same in more accurate models.

\section{Band structures}
\begin{figure}
\begin{center}
\includegraphics[width=\linewidth]{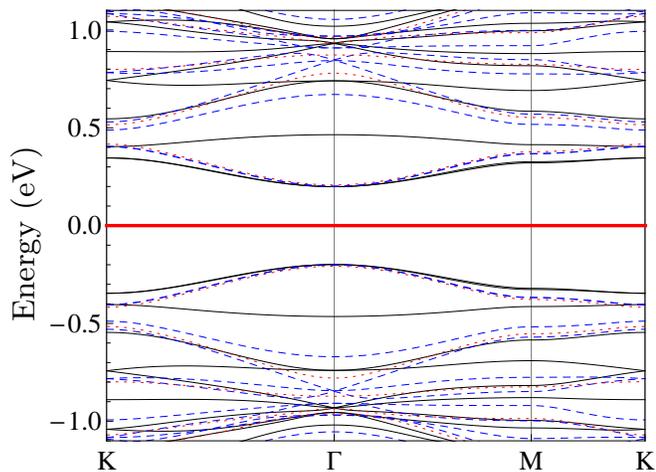}
\caption{(Color online) 
Band structures of the $\{12,5\}$ lattice. 
The solid, black lines show results for perforated graphene, calculated using a TB model.
The blue, dashed (red, dotted) lines correspond to graphene with a periodic mass term
of $V_0=t$, calculated using the TB (DE) model. The thick, red line shows the location of
the Fermi level. Note the perfect electron-hole symmetry in this case, and the agreement
on the magnitude of the band gap between all three methods.
}
\label{fig:bandL12R5Hole}
\end{center}
\end{figure}
In Fig.~\ref{fig:bandL12R5Hole} we show the band structure for a $\{12,5\}$ graphene antidot lattice, i.e.,
periodically perforated graphene, and compare to the case of a periodic mass term, modeled using either the TB
or the DE approach. A sufficiently large mass term
should ensure that electrons are excluded entirely from the region of the mass term, and we thus expect relatively
good correspondence with perforated graphene. In the figure, we consider the case where the mass term is equal in size
to the TB hopping term, $V_0=t$. As expected, we find quite good agreement between all three methods. In particular,
the magnitudes of the band gaps are in near-perfect agreement. Using a finite, but sufficiently large mass term in the DE
model thus yields much better results than models where the limit of infinite mass term is used to impose
boundary conditions on the edge of the hole in the DE model.\cite{Furst2009}
Note that electron-hole symmetry is preserved for all models.
For higher-lying bands, the differences between the DE and TB results
become more pronounced, as the linear approximation of the DE model breaks down. Further, comparing the case of
perforated graphene to that of a periodic mass term in the TB model, we see significant differences in the higher-lying 
bands. However, we note that increasing the mass term further results in excellent agreement with the perforated graphene 
case, for all bands shown.

\begin{figure}
\begin{center}
\includegraphics[width=\linewidth]{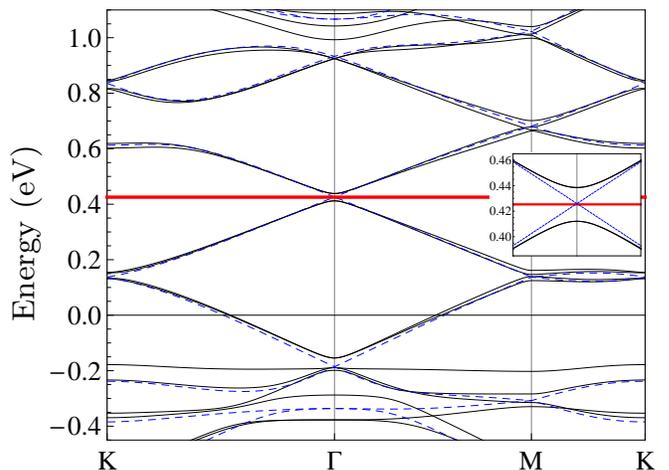}
\caption{(Color online) 
Band structures of a periodically gated $\{12,5\}$ lattice.
The solid, black (blue, dashed) lines show results for periodically gated graphene, calculated using a TB (DE) model.
The gate potential is $V_0=t/2$. The thick, red line shows the location of
the Fermi level. Note the nearly dispersionless band near $-0.2$~eV.
Inset: A zoom of the band structure near the $\Gamma$ point, illustrating the emergence of a band gap
in the TB results and the absence of such a gap in the DE model.
}
\label{fig:bandL12R5V0p5}
\end{center}
\end{figure}
A periodic mass term is expected to induce a gap in graphene due to the fact that it explicitly breaks sublattice symmetry
via the $\hat{\sigma}_z$ operator in the continuum model or, similarly, through the staggered on-site potential in the TB approach.
Contrary to this, analysis of periodic potentials in a DE model of graphene suggests that periodic gating does \emph{not} induce a gap
in graphene around the Fermi level,\cite{Barbier2008, Barbier2009} but rather leads to the generation of new 
Dirac points near the superlattice Brillouin zone 
boundaries.\cite{Park2008} Superlattices lacking inversion symmetry have been suggested as a means of achieving tunable band gaps in
graphene, based on results using a DE model.\cite{Tiwari2009} However, these results were recently found to be based on numerical errors.\cite{Tiwari2012}
Indeed, based on the DE model, a gap cannot be produced by any Hamiltonian that preserves time-reversal symmetry,
i.e. $H=\hat{\sigma}_y H^*\hat{\sigma}_y$, where $\hat{\sigma}_y$ is the Pauli spin matrix while $H^*$ denotes
the complex conjugate (not the Hermitian conjugate) of the Hamiltonian.\cite{Lin2012} 
A pure scalar potential, such as the one 
we consider for periodically gated graphene, see Eq.~(\ref{eq:DE}), preserves this symmetry and the DE model thus suggests 
that periodic gating does \emph{not} open a band gap. Instead, a combination of a scalar as
well as a vector potential is needed.\cite{Lin2012}

In Fig.~\ref{fig:bandL12R5V0p5} we show the band structure of a periodically gated $\{12,5\}$ graphene structure, with a gate
potential of half the TB hopping term, $V_0=t/2$. Results are shown for TB and DE models, respectively.
Contrary to a periodic mass term we see that, as could be expected, periodic gating breaks electron-hole symmetry and 
shifts the Fermi level to higher energies. 
Comparing DE and TB results, we note that there is quite good agreement overall, between the
two methods. However, a crucial difference emerges when considering the bands in close vicinity of the Fermi level, as illustrated
in the inset: while the DE results suggest that periodic gating does not open a band gap, TB results demonstrate that a band gap
\emph{does} occur right at the Fermi level. We attribute this to a local sublattice symmetry breaking at the edge of the gate region
and substantiate this claim below. We note that while a band gap appears, the magnitude of the band gap is of the order of tens
of meV, an order of magnitude smaller than that of the corresponding perforated graphene structure. This dramatic qualitative difference between TB and DE modelling agrees with previous results \cite{Zhang2010} comparing density functional theory and Dirac models for rotated triangular geometries.

\begin{figure}
\begin{center}
\includegraphics[width=\linewidth]{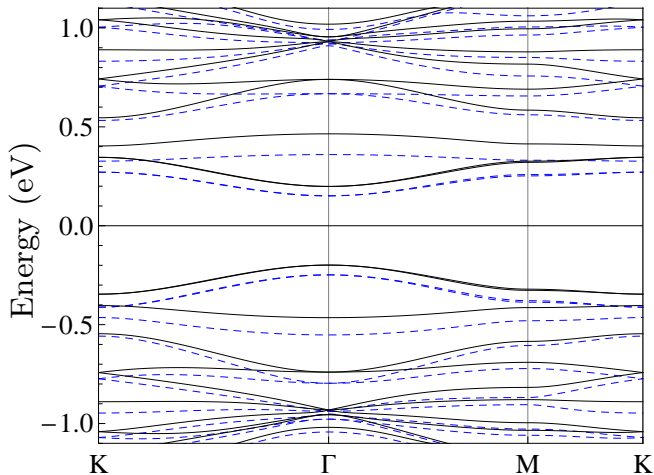}
\caption{(Color online) 
Band structures of a periodically gated $\{12,5\}$ lattice.
The solid, black lines show results for perforated graphene, calculated using a TB model.
The blue, dashed lines correspond to graphene with a periodic gate potential
of $V_0=10 t$, calculated using the TB model. Bands near the original Dirac energy of graphene are shown. 
For the gated structure, the Fermi level is far removed from the Dirac energy of 
graphene, outside the range of the figure, and no band gap occurs  at the Fermi level for this structure.
}
\label{fig:bandL12R5V10}
\end{center}
\end{figure}
Above, we illustrated how a sufficiently large mass term serves as an excellent model of a hole in graphene, 
see Fig.~\ref{fig:bandL12R5Hole}. Because a simple scalar potential cannot confine Dirac electrons \cite{Novoselov2006,Beenakker2008}, 
one would expect that modeling the hole via a large gate potential would be inaccurate. In Fig.~\ref{fig:bandL12R5V10}
we show the band structure of periodically gated graphene, with a very large gate potential of 
$V_0 = 10t$.\cite{footnote:V10} 
For comparison, we also show the corresponding perforated graphene structure. Contrary to the aforementioned expectations, 
we see that the periodically gated graphene structure is an excellent model of perforated graphene. We note that increasing 
the gate potential further results in near-perfect agreement between the periodically gated and the perforated structures. 
With a gate potential of $V = 10t$ we are way beyond the linear regime of the band structure, for which a Dirac treatment of graphene is viable, 
which explains why the theoretical arguments pertaining to Dirac electrons break down in this case. 

\subsection{Gate region overlap}
\begin{figure}
\begin{center}
\includegraphics[width=\linewidth]{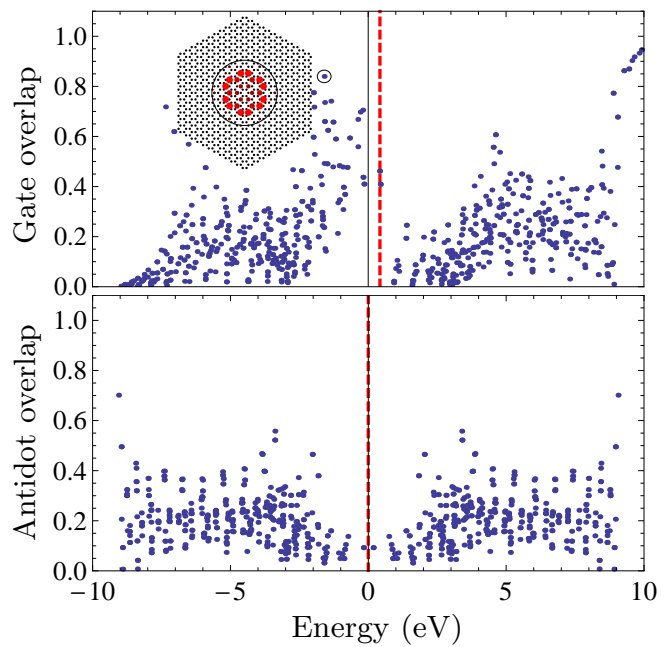}
\caption{(Color online)
Overlap of eigenstates with the gate region, calculated at the $\Gamma$ point for the $\{12,5\}$ lattice 
with a gate potential (upper panel) or mass term (lower panel) of $V_0=t/2$.
The Fermi level is indicated by the dashed, vertical line.
The inset in the upper panel shows the eigenstate corresponding to the state highlighted with a circle. 
The size of the filled, colored circles indicates the absolute square of the wavefunction.
The black circle indicates the radius of the gate region.
}
\label{fig:ADoverlap}
\end{center}
\end{figure}
Returning now to the band structure for the periodically gated $\{12,5\}$ lattice, shown in Fig.~\ref{fig:bandL12R5V0p5},
we note the appearance of a nearly dispersionless band near $-0.2$~eV. This state is
localized predominantly within the gate region. In the upper panel of Fig.~\ref{fig:ADoverlap} we show the overlap of 
all eigenstates with the 
gate region as a function of energy, calculated at the $\Gamma$ point. For comparison, we show the corresponding results
for a periodic mass term in the lower panel.
We note that several states exist, which have significant
overlap with the gate region, also at energies below the Fermi level. An example of one such state is shown in the figure.
As the gate potential is increased further, these states become less energetically favorable, and are eventually all situated 
at energies well above the Fermi level. In stark contrast to this, a periodic mass term dictates perfect electron-hole symmetry,
and thus always predicts states below the Fermi level having significant overlap with the gate region. In fact, as the mass term
is increased, states nearly entirely localized within the mass term region develop at both extrema of the spectrum.
Below, we will illustrate how this fundamental difference between a mass term and a scalar potential manifests itself via
the dependence of the band gap on the magnitude of the gate potential for periodically gated graphene.

\section{Band gaps in periodically gated graphene}
\begin{figure}
\begin{center}
\includegraphics[width=\linewidth]{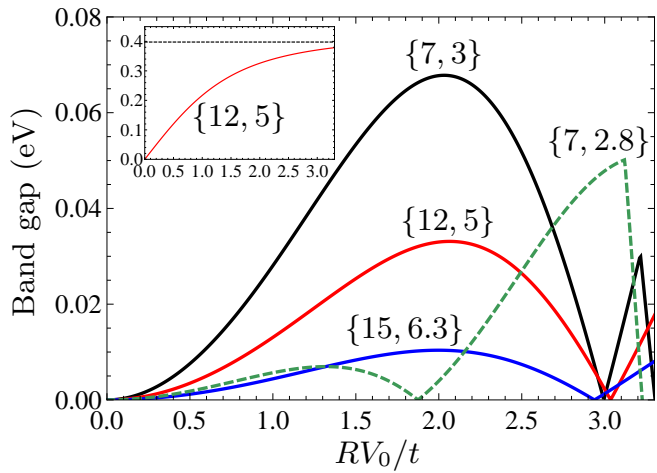}
\caption{(Color online)
Band gap at the Fermi level for periodically gate graphene, as a function of the
gated radius (in units of the graphene lattice constant) times the gate potential in units of the TB hopping term. Results are shown for three
different lattices (solid lines), with roughly equal ratios $R/L$ of the radius of the gate region to the side length of
the hexagonal unit cell. The dashed line shows the results for the $\{7,2.8\}$ lattice, which has roughly the same
$R/L$ ratio. Inset: Results for $\{12,5\}$, 
when the potential is replaced by a mass term. The dashed line indicates the band gap for a perforated graphene structure.
}
\label{fig:gapVsGate}
\end{center}
\end{figure}
Having determined that a TB treatment of periodically gated graphene does indeed suggest the opening of a band gap
at the Fermi level, we now proceed to investigate the behavior of the band gap magnitude in more detail. From hereon, all
results shown have been calculated using the TB model. 

In Fig.~\ref{fig:gapVsGate}, the solid lines show the magnitude of the band gap at the Fermi level for three different lattices, 
$\{7,3\}$, $\{12,5\}$, and $\{15, 6.3\}$, all of which have approximately
the same ratio $R/L$ of gate radius to side length of the hexagonal unit cell. When plotted against the
gate radius times the gate potential, the resulting curves emerge as simple scaled versions of each other,
as seen in Fig.~\ref{fig:gapVsGate}. While, initially, raising the gate potential increases the band gap,
a maximum gap is reached at a certain gate potential, after which the band gap 
diminishes. This behavior is completely different from the case where the potential is replaced
by a mass term, as illustrated in the inset of the figure. In this case, the band gap continues to increase until
a saturation point is reached in the limit where the structure resembles that of perforated graphene. 
While the three periodic lattices indicated with solid lines in Fig.~\ref{fig:gapVsGate} result in similar
dependencies of the gap on $RV_0$, we stress that this is \emph{not} the case for all lattices, even if the ratio $R/L$ is 
approximately the same. To illustrate this, we also show in Fig.~\ref{fig:gapVsGate} results for the 
$\{7, 2.8\}$ lattice. The dependence of the band gap on gate potential differs markedly for this lattice.
This indicates that the exact geometry at the edge of the gate region plays a large role in determining 
the band gap, in agreement with findings in Ref.~\onlinecite{Zhang2010}.

\subsection{Edge dependence}
\begin{figure}
\begin{center}
\includegraphics[width=\linewidth]{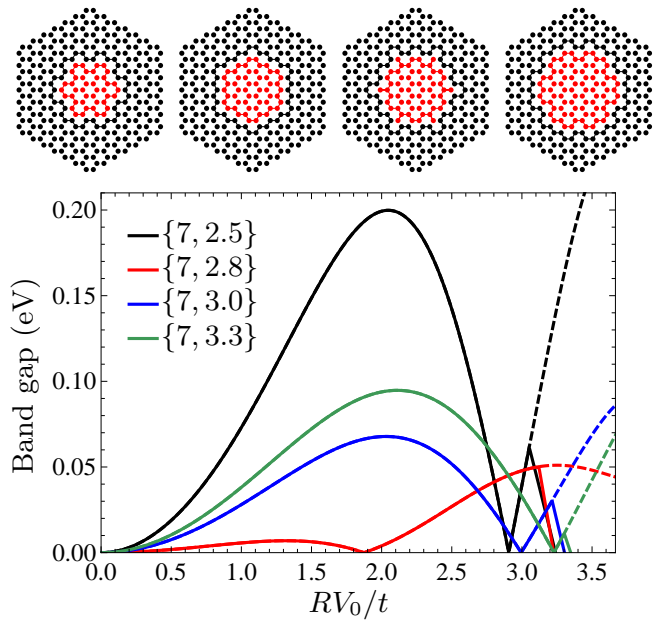}
\caption{(Color online)
Band gap at the Fermi level for periodically gated graphene. The band gap is shown as a function of the
gate radius (in units of the graphene lattice constant) times the gate potential in units of the TB hopping term. 
Results are shown for lattices $\{7,R\}$ with varying $R$. 
Below $RV_0\simeq 3t$ all gaps are direct ($\Gamma$--$\Gamma$). Above this transition the $\Gamma$--$\Gamma$ gaps
(dashed lines) exceed the indirect $\Gamma$--$K$ gap.
The unit cells of the $\{7,R\}$ lattices are shown above, in order of increasing
radius. The edge geometry is highlighted.
}
\label{fig:gapVsGateL7}
\end{center}
\end{figure}
To illustrate in more detail the role of the edge in determining the band gap, we show in Fig.~\ref{fig:gapVsGateL7} the band gap
as a function of the gate potential, for lattices $\{7,R\}$ with increasing values of $R$. The radius is increased in the minimum
steps resulting in new geometries. The structures with $R\in\{2.5,\, 3.0,\, 3.3\}$ show quite similar behaviors. In particular, a maximum
band gap is reached at $RV_0\simeq 2t$ in all three cases. The band gap then closes, but reopens once more as the gate potential is
increased further. 
Around $RV_0\simeq 3t$ the band gap changes from direct ($\Gamma$--$\Gamma$) to indirect ($\Gamma$--$K$) as the gate voltage is raised.
The dashed lines in the figure illustrate the $\Gamma$--$\Gamma$ gap above the direct to indirect-gap transition.
However, after a slight further increase of the gate voltage,
the final closing of the band gap occurs as the energy at the $K$ point moves below that at the $\Gamma$
point, resulting in crossing bands at the Fermi level. 
Finally, we note that while the three lattices show similar behavior, the dependence of the 
band gap on the radius of the gate region is clearly not monotonic, and a larger gate region does not necessarily result 
in a larger band gap.

In contrast to the similarities of the other three structures, the dependence of the band gap on 
the gate potential for the $\{7,2.8\}$ lattice differs greatly. In the upper panel of Fig.~\ref{fig:gapVsGateL7} we show the unit cells 
corresponding to the $\{7,R\}$ lattices considered, with the edge geometries highlighted. 
The $\{7,2.8\}$ lattice stands out from the rest of the geometries, in that the entire edge of the gate region is made
up from several pure zigzag edges. We stress that the sublattice imbalance for the entire edge is zero, while there is a
local sublattice imbalance on the individual straight zigzag edges. In contrast to this, the other geometries have 
gate regions with zigzag as well as armchair edges. We find that the general trend is for zigzag edges to 
quench the band gap of the periodically gated graphene structures, which we have also verified via calculations of 
gate regions of hexagonal symmetry, which always have pure zigzag edges. This trend can be explained by noting that 
pure zigzag edges, such as, e.g., in zigzag graphene nanoribbons \cite{Nakada1996,Brey2006} or graphene antidot lattices with
triangular holes \cite{Vanevic2009,Furst2009a,Gunst2011}, lead to localized midgap states.\cite{Inui1994} 
For periodically gated graphene
the edge is defined by a finite potential, rather than a complete absence of carbon atoms, so we expect the tendency of 
electrons to localize on the edge to be less pronounced. Nevertheless, our findings suggest that local zigzag geometry still
has the effect of quenching the band gap. Since, in general, larger circular
holes will have longer regions of zigzag geometry at the edge of the gate region, this explains why larger gate
regions will not invariably lead to larger band gaps. In the present case, we note that the $\{7,3.3\}$ structure indeed
has a significantly smaller band gap than the $\{7, 2.5\}$ structure. The $\{7,3.0\}$ lattice is unique in that
the equivalent of dangling bonds are present at the edge of the gate region, which further decrease the magnitude
of the band gap.

\subsection{Dependence on gate region overlap}
\begin{figure}
\begin{center}
\includegraphics[width=\linewidth]{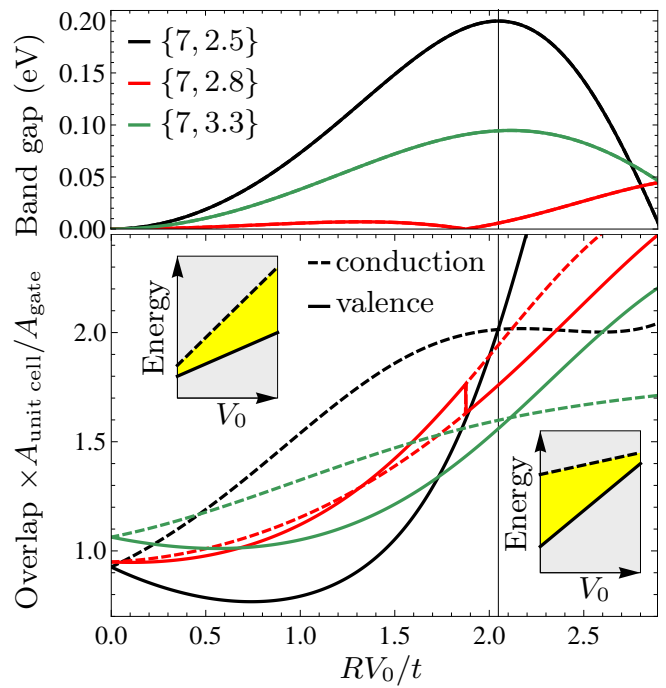}
\caption{(Color online) Overlap of the eigenstates nearest the Fermi level as a function of the
gate radius (in units of the graphene lattice constant) times the gate potential in units of the TB hopping term. 
The solid lines show the overlap of the highest valence band state, while the dashed lines
show the overlap of the lowest conduction band state. The overlap is shown relative to the ratio 
between the area of the gate region, and the unit cell area.
The upper panel repeats the data from Fig.~\ref{fig:gapVsGateL7} showing the band gap.
Note that the overlaps of the two states are equal exactly when the band gap is at a maximum, as highlighted
for the $\{7,2.5\}$ lattice with the vertical black line.
The left (right) inset illustrates schematically the dependence of the conduction and valence band edges on the gate 
potential, in the regime where the overlap with the gate of the state at the valence band edge is smaller (larger) than
that of the state at the conductance band edge.
}
\label{fig:OLvsV}
\end{center}
\end{figure}
First-order perturbation theory suggests that the dependence of the energy of the eigenstate on the gate
potential be proportional to the overlap of the state with the gate region, i.e.,
$\partial E/\partial V_0\propto \int_\mathrm{gate} d\mathbf{r} |\Psi(\mathbf{r})|^2$. We thus expect the overlap
with the gate region of the two eigenstates closest to the Fermi level to be a crucial parameter in describing
the opening and quenching of the band gap as the gate voltage is varied. We will also see that it illustrates the
crucial differences between graphene and ordinary two-dimensional electron gases.
In Fig.~\ref{fig:OLvsV} we show the overlap of the eigenstate with the gate region as a function
of the magnitude of the potential. The overlap is shown for the eigenstates at the valence and conduction band edges, and 
normalized by the ratio between gate and unit cell areas. A value of one thus indicates that
the overlap with the gate region is the same as if the eigenstate is evenly distributed throughout the unit cell, while
a value larger (smaller) than one suggests that the eigenstate is localized predominantly inside (outside) the gate 
region. As we saw also in Fig.~\ref{fig:ADoverlap}, the states near the Fermi level both have quite large overlaps with
the gate region, even when the potential is of the order of the TB hopping term. Initially, for low values of the
gate potential, the overlap with the gate region of the unoccupied state in the conduction band is larger than the corresponding
overlap of the occupied state in the valence band. Relying on first-order perturbation theory we thus expect the
energy of the conductance band state to increase more strongly with the gate potential than the valence band state,
resulting in a larger band gap as the gate potential is raised, as illustrated in the left inset of Fig.~\ref{fig:OLvsV}.
However, contrary to what would be expected for
an ordinary two-dimensional electron gas, we see that as the potential is increased further, the valence band state 
also becomes localized predominantly within the gate region. Indeed, eventually the overlap of the valence band state
with the gate region becomes larger than the one of the conduction band state, which results in a quenching of the band gap
as the potential is increased further, as illustrated in the right inset of Fig.~\ref{fig:OLvsV}. 
We note that the point where the overlap of the two states with the gate region
become equal exactly matches the point where the band gap is at a maximum. This is illustrated by the vertical, black line in
the figure.
The strong influence of the exact edge geometry is apparent,
manifesting itself in a qualitatively different dependence of the overlap on gate voltage for the $\{7,2.8\}$ lattice. 
In particular, while the gate region overlap of the valence band state of the
$\{7,2.5\}$ and $\{7,3.3\}$ lattices initially decreases with the size of the potential, both valence and conduction band
states immediately start localizing within the gate region for the $\{7,2.8\}$ structure. This leads to much faster
quenching of the initial band gap.

\subsection{Realistic potential profiles}
As we have illustrated above, the band gap of periodically gated graphene depends strongly on the edge geometry 
at the boundary between the gated and the non-gated region. So far, we have used a simple step function to model the
spatial dependence of the potential due to the gate. However, it is obvious that in actual realizations of periodically
gated graphene, some form of smoothing of the potential will inevitably be present. Due to the intricate relationship between 
the band gap and the edge geometry, it is relevant to investigate the effect of smoothing out the potential. 
In particular, since
the DE model predicts no gap at all, one may wonder whether smoothing will cause the gap to close entirely. Previous studies
have included smoothing of the gate potential, but with a smearing distance of the order
$0.1$~\AA,\cite{Zhang2010} small enough that an atomically resolved edge can still be defined.

To model a more realistic gate potential, we use an analytical expression for the potential distribution
resulting from a constant potential disk in an insulating plane, obtained by direct solution of the Laplace equation. 
In cylindrical coordinates, this reads as\cite{Nanis1979}
\begin{equation}
V(r,z) = 
\frac{2V_0}{\pi}\sin^{-1}\!\!\left(\frac{2R}{\sqrt{\left(r-R\right)^2+z^2}+\sqrt{\left(r+R\right)^2+z^2}}\right),
\label{eq:potential}
\end{equation}
with $z$ the distance above the gate, while $r$ is the distance from the center of the disk. 
Note that for $z=0$ the expression simplifies to
$V(r,0) =  V_0$ for $r\leq R$ while $V(r,0) = 2V_0 \sin^{-1}(R/r)/\pi$ for $r>R$. Of course, more exact approaches
such as, e.g., finite-element methods, could be used to determine the potential distribution from a realistic back
gate. However, we choose to use this relatively simple analytical expression, since we are mainly interested in discussing
the general trends that occur as the edge of the potential region becomes less well-defined. One could imagine
more elaborate setups that would generate sharper potential distributions. To include such possibilities, 
we consider a modified potential distribution $\tilde{V}(r,z) = V_0 [V(r,z)/V_0]^\eta$,
with the additional parameter $\eta$, which allows us to control the smoothing of the potential further. 
As $\eta\rightarrow\infty$ we
approach the limit where the potential is described by a Heaviside step function, as in the results presented so far.
We note that Eq.~(\ref{eq:potential}) is derived for an isolated constant potential disk rather than a periodic array of gates.
Ignoring coupling between the different gates, one simple way of improving this model would be to add the potentials 
generated from the nearest-neighbor gates, to 
account for the overlap between them. However, this would merely serve to smoothen out the potential further, as well as
add a constant background potential, effectively decreasing the height of the potential barrier. 
Here, we are interested in illustrating the critical dependence of the band gap
on smoothing out the potential, so we are adopting a `best case' scenario, which also means that we will use
$z=0$ throughout, assuming that the graphene layer is deposited directly on the periodic gates, with no insulating layer
in-between.

\begin{figure}
\begin{center}
\includegraphics[width=\linewidth]{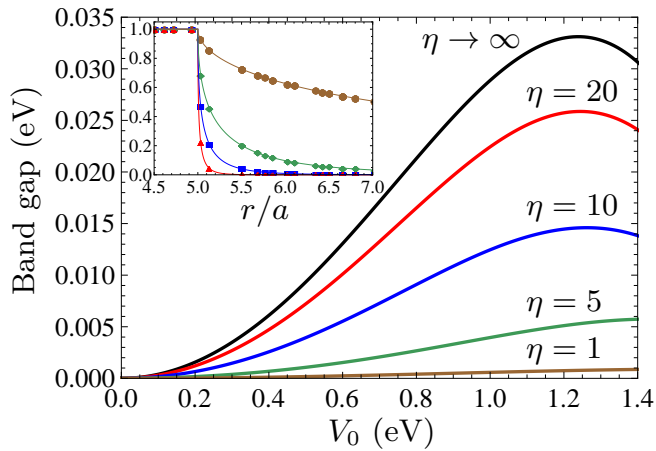}
\caption{(Color online)
Band gap as a function of gate potential for the $\{12,5\}$ periodically gated graphene lattice.
The potential distribution due to the periodic gates is modelled via Eq.~(\ref{eq:potential}).
We assume the distance from the plane of the gate to the graphene layer
is zero.  Results are shown for increased values of the $\eta$ parameter, which determines the strength
of the smoothing. The inset illustrates the potential distribution $\tilde{V}/V_0$ for each case, with markers
indicating the radial position of the carbon atoms.
}
\label{fig:GapvsGateSmooth}
\end{center}
\end{figure}
In Fig.~\ref{fig:GapvsGateSmooth} we show the band gap for a $\{12,5\}$ lattice as a function of gate
potential, for increased values of the $\eta$ parameter. While for $\eta\rightarrow\infty$, corresponding
to a Heaviside step function distribution, the maximum band gap is about $33$~meV, the band gap for
$\eta=1$ is drastically lower, with a maximum value of only $0.9$~meV. As we artificially decrease
the amount of smoothing by raising the value of $\eta$, we slowly recover the maximum band gap
attainable. However, we stress that even for $\eta=20$, which as shown in the inset of the figure
amounts to smoothing over a range of little more than a single carbon atom, the maximum band gap has decreased by
more than 20\% from the value at $\eta\rightarrow\infty$. This suggests that the band gap does indeed critically 
depend on an edge effect, which is very quickly washed out as the potential step is smoothed out over several carbon
atoms. This is in agreement with previous studies, which have indicated that intervalley scattering is crucial
in describing the band gap of periodically gated graphene.\cite{Zhang2010} In order for a scalar potential to
induce intervalley scattering, it must vary significantly on a scale of the carbon-carbon distance, so that
a local sublattice asymmetry is introduced.

\section{Summary}
By employing a tight-binding description of graphene, we have shown that, contrary to what is predicted on basis
of a continuum model, it is indeed possible to induce a band gap in graphene via periodic, electrostatic gating.
Further, if the magnitude of the potential is made sufficiently large, periodically gated graphene is an accurate model
for perforated graphene structures, with one caveat, namely that the Fermi level is far removed from the location of the
band gap. For smaller, more realistic values of the gate potential, a band gap appears right at the Fermi level.
However, we find that the band gap is orders of magnitude smaller than that of the corresponding perforated
graphene structure.

The dependence of the band gap on the gate potential is highly non-trivial, and entirely different from the case where
graphene is modulated by a periodic mass term. In particular, a maximum magnitude of the band gap is reached, after which
increasing the gate potential further quenches the gap. Also, a transition from a direct ($\Gamma$--$\Gamma$) 
to an indirect ($\Gamma$--$K$)
semiconductor occurs for larger gate potentials. The exact magnitude and dependence of the band gap on gate potential depends
critically on the precise geometry of the edge of the gate region. In particular, large regions of local zigzag geometries 
tend to result in significantly smaller band gaps than geometries where armchair edges dominate.

Because the emergence of a band gap relies on a local sublattice asymmetry, we find that it is extremely fragile.
If smoothing is introduced in the potential distribution, such that the edge of the gate region is no longer atomically resolved,
the magnitude of the band gap drops significantly. Even if the smoothing occurs over a range of little more than a single
carbon atom, we find that the maximum band gap decreases to less than 80\% of the value for a perfectly defined edge. 
This presents a serious challenge to opening a band gap in graphene via periodic gating.

\begin{acknowledgments}
The work by J.G.P. is financially supported by the Danish Council for Independent Research, FTP grant numbers
11-105204 and 11-120941. The Center for Nanostructured Graphene (CNG) is sponsored by the Danish National 
Research Foundation. We thank Prof. Antti-Pekka Jauho for helpful comments during the development of the manuscript.
\end{acknowledgments}


\begin{thebibliography}{33}
\bibitem[{foot()}]{footnote:V10}
\bibinfo{note}{While this is a very large value of the gate potential, our point here is not whether such
a periodic gate can be realized, but rather that there is theoretical agreement between this model and that of perforated
graphene.}
\end{thebibliography}
\end{document}